\documentstyle[aps]{revtex}

\begin{document}
\title{Variational Principle for Relativistic Fluid Dynamics}
\author{Hans-Thomas Elze$^{1}$, Yogiro Hama$^{2}$, Takeshi Kodama$^{1}$, Mart\'{\i}n
Makler$^{3}$ and Johann Rafelski$^{4}$}
\address{$^{1}$Instituto de F\'{\i}sica, Universidade Federal do Rio de Janeiro, \\
CP 68.528, 21945-970 Rio de Janeiro, Brasil\\
$^{2}$Instituto de F\'{\i}sica, Universidade de S\~{a}o Paulo, CP~66318,\\
05389-970 Sao Paulo, Brasil\\
$^{3}$Centro Brasileiro de Pesquisas F\'{\i}sicas, Rua Xavier Sigaud 150,\\
22290-160 Rio de Janeiro Brasil\\
$^{4}$Physics Department, University of Arizona, Tucson, AZ 85721 USA}
\maketitle

\begin{abstract}
The variational principle for the special and general relativistic
hydrodynamics are discussed in view of its application to obtain approximate
solutions to these problems. We show that effective Lagrangians can be
obtained for suitable ansatz for the dynamical variables such as density
profile of the system. As an example, the relativistic version of spherical
droplet motion (Rayleigh-Plesset equation) is derived from a simple
Lagrangian. For the general relativistic case the most general Lagrangian
for spherically symmetric systems is given.
\end{abstract}

\section{Introduction}

First applications of relativistic hydrodynamics to the process of
multiparticle production in high-energy hadronic collisions can be found in
the works of Fermi and Landau in the early 1950's \cite{Fermi,Landau}.
Recently, extensive studies of the relativistic motion of fluids have been
done with respect to the analysis of relativistic heavy-ion collision
processes\cite{Stocker,Strottman,Csernai}. In fact, a hydrodynamic
description of high-energy hadronic and nuclear collisions has been
successful in reproducing global features of these processes, such as
multiplicity and transverse energy distributions. From the theoretical point
of view, however, the foundation of the hydrodynamical picture for these
processes is not a trivial matter. This is because the use of hydrodynamic
equations of motion assumes implicitly the local thermal equilibrium via an
equation of state of the matter. This means that the relaxation time scale
and the mean free path should be much smaller compared to, respectively, the
hydrodynamical time scale and spatial size of the system. In this sense, one
may wonder whether these conditions could easily be met for hadronic and
nuclear collisions (for the collision of heavier nuclei they are expected to
be approximately fulfilled for some specific scenario).

On the other hand, from the kinematical point of view, apart from the use of
the equation of state, the equations of hydrodynamics are nothing but the
conservation laws of energy and momentum, together with other conserved
quantities such as charge. In this sense, for any process where the dynamics
of flow is an important factor, a hydrodynamic framework should be a natural
first step, at least at the level of phenomenology. The effects of finite
relaxation time and mean-free path might be implemented at a later stage by
using an effective equation of state, incorporating viscosity and heat
conductivity, or some simplified transport equations, see Ref. \cite{Elze}
and references therein.

Another important arena of extensive application of relativistic
hydrodynamics is found in cosmology and high-energy astrophysics, such as
the gravitational collapse of a stellar core to form a neutron star or a
black hole, relativistic blast waves for the models of gamma ray bursts, etc.
\cite{Cosmos,Astro1,Astro2,GB}. In these cases, the assumption of the local
thermodynamical equilibrium is considered to be well justified. However, in
the astrophysical applications we have not only to face the large scale
systems but also to deal with the long range gravitational field
simultaneously. For these reasons, the computer simulations of
hydrodynamical scenarios for astrophysical problems usually become extremely
expensive.

The relativistic hydrodynamics is a {\it local} description of the
conservation laws, written in terms of the energy-momentum tensor as 
\begin{equation}
\partial _{\mu }T^{\mu \nu }=0.  \label{divT}
\end{equation}
This is a set of coupled partial differential equations which, in general,
are difficult to solve exactly. Except for a few analytical solutions known
for special cases, we have usually to resort to numerical solutions even for
a simplest geometry, like one-dimensional or spherically symmetric cases. In
the most of the cases, the numerical approach together with a realistic
equation of state becomes prohibitively expensive, especially when coupled
to some transport equations such as those for neutrinos in the case of
stellar collapse or supernova explosions\cite{Mezza}.

In addition to the difficulties of solving the hydrodynamical equations,
frequently we encounter with the situation where even the equation of state
of the matter is not known precisely. Rather, we apply the hydrodynamical
models to infer the properties of the matter involved in the process. In
such cases we do not need the very precise local features of the
hydrodynamical motion (for example, sound ripples, small local
perturbations, etc.) but rather the global flow motion which characterizes
the dynamics of the system assuming a given equation of state.

For the reasons cited above, in spite of the presence of highly
sophisticated techniques for hydrodynamic numerical calculations, some
problems require rather simpler approaches which allow to analyze the
dynamics of the system more effectively. In such cases, extremely local
properties should be smeared out effectively, in order to extract global
features of the flow more directly. As an example, we know that some global
features of the high-energy hadronic and nuclear collisions can already be
described by a simple fireball model\cite{fball}. Presently we aim at a
dynamical scheme which improves the simplest fireball model in the direction
of a more complete hydrodynamical description.

We here introduce the method of effective action based on the variational
principle to hydrodynamic equations of motion. As is well-known, the
variational approach has practical advantages besides its formal side. Once
the variational principle is established, we can use the method to obtain
the optimal parameters of a given family of trial solutions. The effective
Lagrangian and variational approach\cite{Leff,Retard}, introduced to
incorporate the effect of local turbulent motion in an effective way into a
supernova explosion mechanism, is such an example. It was shown that such an
approach is also useful to discuss the dynamics of a sonoluminescencing
bubble in a fluid\cite{RP,KTR}. There, the effective Lagrangian method was
shown to be very useful in generalizing the so-called Rayleigh-Plesset
equation to include in a simple way the effects of gas dynamics inside the
bubble. There exist some analogous problems to the dynamics of a
sonoluminescence bubble in the domain of relativistic energies, such as QGP
or astrophysical fireballs. Thus, the relativistic generalization of the
classical Rayleigh-Plesset equation will be useful.

In the present work, we generalize the effective Lagrangian method to the
relativistic hydrodynamics. By suitable parametrizations of the density
profile of the system, approximate but very simple solutions of relativistic
hydrodynamical models can be derived in this approach. In particular, we
derive a relativistic generalization of the Rayleigh-Plesset equation and
discuss the effect of relativity for the homologous motion of gas and fluid.

Frequently the flow of the matter accompanies the production of entropy. In
particular, when a shock wave is generated, the violent dynamical change of
the density leads to a highly turbulent regime, which cascades into a
smaller scale complex fluid motion and ultimately thermalize. In order to
incorporate such effects of non-adiabatic processes and simulate the
dynamics of shock wave as a thin domain of non-adiabatic flow, Neumann and
Richtmyer\cite{Neuman} introduced the method of the pseudo-viscosity which
is still used extensively in many areas. We show that this approach can well
be incorporated in our formalism and consequently, the relativistic
generalization of the pseudo-viscosity method is easily obtained in our
context.

In the astrophysical applications of relativistic hydrodynamics, the
inclusion of the gravitational field is essential. It has been discussed by
several authors that the general relativistic hydrodynamical equations can
also be derived from the action principle\cite{Taub,Schutz,Ray,Bogos}. In
this paper, we derive a simple general relativistic effective Lagrangian for
spherically symmetric systems and deduce explicitly from it the equation of
motion of Misner and Sharp\cite{MS} for gravitationally collapsing object.
We also show that the concept of general coordinate system allows us to use
a comoving Lagrangian frame in obtaining the effective Lagrangian of the
special relativistic hydrodynamics.

We organize this paper as follows. In Sec. II, we first review the
variational formulation of the relativistic hydrodynamics. Then, in Sec.III,
we apply the variational scheme to spherically symmetric cases and establish
an effective Lagrangian for the variational parameters of the density
profile function. In the case of a homogeneous gas bubble in an infinite
fluid, this equation is a relativistic generalization of the
Rayleigh-Plesset equation well-known for fluid acoustic theories. In Sec.
IV, we discuss nonadiabatic process and generalize the pseudo-viscosity of
Neumann and Richtmeyer in the context of our relativistic variational
principle. In Sec.V, we extend our approach to the general relativistic
case, where, the metric functions are chosen as the dynamical variables. In
the case of spherically symmetric system, we derive explicitly the effective
Lagrangian for the comoving coordinate system. We discuss the relation
between the comoving frame and space-fixed coordinate system. In Sec.VI, we
summarize our present work.

\section{Variational Approach}

Although not commonly found in general textbooks, the variational
formulation of hydrodynamics has been studied by several authors\cite
{Taub,Schutz,Ray,Bogos,Yourgrau,Herivel,Lin,Chiueh}. For the sake of later
discussion, let us first review briefly how the relativistic hydrodynamical
equations of motion are derived from a variational principle. In the
following, we take the velocity of light is unity, $c=1$. Let the velocity
field of the matter be 
\begin{equation}
\vec{v}=\vec{v}(\vec{r},t).  \label{v}
\end{equation}
In order to keep the manifestly covariant notation, we express the flow in
terms of a four-vector, $u^{\mu }(x)$, where 
\begin{equation}
u^{0}=\gamma ,\;\;\vec{u}\;=\gamma \vec{v}.  \label{umu}
\end{equation}
and 
\begin{equation}
u_{\mu }u^{\mu }=1.  \label{uu=1}
\end{equation}
The flow of matter induces a change in the specific volume occupied by the
matter. In order to facilitate the following discussion, we consider the
case where there exists some conserved quantity, say the baryon number. Let
the local density of this conserved quantity in the comoving frame be $n$.
Then we have 
\begin{equation}
\partial _{\mu }\left( nu^{\mu }\right) =0.  \label{conti}
\end{equation}
We also define the specific volume $V$ as 
\begin{equation}
V=\frac{1}{n}.  \label{V}
\end{equation}
Let us write the energy of the matter in this volume as 
\begin{equation}
E=\varepsilon V,
\end{equation}
where $\varepsilon $ is the energy density. The assumption of local
equilibrium leads to the validity of the thermodynamical relations, such as 
\begin{equation}
\left( \frac{\partial E}{\partial V}\right) _{S}=-P,  \label{p}
\end{equation}
where $S$ is the entropy of the matter in the volume and $P$ the pressure.
In terms of the energy density this implies 
\begin{equation}
\left( \frac{\partial \varepsilon }{\partial n}\right) _{S}\;=\frac{%
\varepsilon +P}{n}.  \label{dedv}
\end{equation}

Now take the action, 
\begin{equation}
I_{M}=\int d^{4}x\left\{ -\varepsilon (n)+\xi (x)\partial _{\mu }\left(
nu^{\mu }\right) +\;\frac{1}{2}\zeta (x)\left( u^{\mu }u_{\mu }-1\right)
\right\} .  \label{L}
\end{equation}
and state the variational principle as 
\[
\delta I_{M}=0, 
\]
for arbitrary variations in $u^{\mu },n,\xi $, and $\zeta $. Then, as we
will show in the following, Euler's equation for the relativistic fluid
motion can be derived formally from the variational principle\footnote{%
In fact, the action above only applies to the case of non-rotational flow.
It is also possible to formulate the variational principle for more general
flow pattern. See \cite{Schutz,Ray} for discussion.}. Note that the last two
terms in Eq.(\ref{L}) represent the constraints among variables $u^{\mu }$
and $n$. As we see in the next section, for the practical usage of this
variational approach, it is convenient to choose the parametrization of $%
u^{\mu }$ and $n$ in such a way that the constraints are automatically
satisfied so that the Lagrangian multipliers $\xi $ and $\zeta $ do not
enter into the calculation.

The variations in $\xi $ and $\zeta $ lead immediately to the constraints,
Eqs.(\ref{uu=1}) and (\ref{conti}). Applying an integration by parts to the
second term in Eq.(\ref{L}), the action can also be written as 
\begin{equation}
I_{M}=\int d^{4}x\left\{ -\varepsilon (n)-nu^{\mu }\partial _{\mu }\xi
(x)\;+\;\frac{1}{2}\zeta (x)\left( u^{\mu }u_{\mu }-1\right) \right\} .
\label{L1}
\end{equation}
We remark that the derivation in this section is equally valid in general
coordinate systems. In this case, the partial derivative, $\partial ^{\mu },$
in Eq.(\ref{L1}) should be replaced by the appropriate covariant derivative,
and correspondingly in the following equations. Furthermore, the volume
element $d^{4}x$ should be replaced by the invariant volume element $\sqrt{-g%
}d^{4}x$\cite{Weinberg}.

The variation with respect to $n$ leads to 
\begin{equation}
-\frac{\delta \varepsilon }{\delta n}-u^{\mu }\partial _{\mu }\xi =0.
\label{a}
\end{equation}
Note that, if the motion of the fluid is not adiabatic, then $\delta
\varepsilon /\delta n$ does not necessarily be equal to the usual derivative 
$d\varepsilon /dn$ (see the later discussion). On the other hand, the
variation in $u^{\mu }$ leads to 
\begin{equation}
\partial _{\mu }\xi =\frac{1}{n}\zeta u_{\mu },  \label{c}
\end{equation}
and substituting this into Eq.(\ref{a}), we obtain: 
\begin{equation}
\zeta =-n\frac{\delta \varepsilon }{\delta n},  \label{zeta}
\end{equation}
where we have used $u^{2}=1$, cf. Eq.(\ref{uu=1}). Thus, Eq.(\ref{c}) is
rewritten as 
\begin{equation}
\partial _{\mu }\xi =-\frac{\delta \varepsilon }{\delta n}u_{\mu }.
\label{divl}
\end{equation}
Taking the contraction of both sides with the four-velocity, we have 
\begin{equation}
u^{\mu }\partial _{\mu }\xi =-\frac{\delta \varepsilon }{\delta n}u_{\mu
}u^{\mu }=-\frac{\delta \varepsilon }{\delta n}.
\end{equation}
Thus we conclude that 
\begin{equation}
\frac{\partial \xi }{\partial \tau }=-\frac{\delta \varepsilon }{\delta n},
\label{dldtau}
\end{equation}
since $u^{\mu }\partial _{\mu }\xi =\partial \xi /\partial \tau $.

In the usual hydrodynamic equations, we assume that the matter is always in
thermodynamical equilibrium. Furthermore, if there is no viscosity or heat
conduction, the energy change associated to the motion is adiabatic, that
is, the change in specific energy $E$ caused by the change in the specific
volume $V$ is given by 
\begin{equation}
\delta E=-P\delta V,  \label{dE=pdV}
\end{equation}
where $V=1/n$ and $P$ is the pressure. In such cases, we have 
\begin{equation}
\frac{\partial \varepsilon }{\partial n}\rightarrow \frac{\delta (nE)}{%
\delta n}=\frac{\varepsilon +P}{n},  \label{adiab1}
\end{equation}
and 
\begin{equation}
d\left( \frac{\delta \varepsilon }{\delta n}\right) \rightarrow \frac{1}{n}%
dP.  \label{adiab2}
\end{equation}
Therefore, for adiabatic changes of the density $n,$ Eq.(\ref{dldtau})
becomes 
\begin{equation}
\partial _{\mu }\frac{\partial \xi }{\partial \tau }=-\frac{1}{n}\partial
_{\mu }P.  \label{gradp}
\end{equation}
On the other hand, we have also 
\begin{eqnarray}
\partial _{\mu }\frac{\partial \xi }{\partial \tau } &=&\partial _{\mu
}(u^{\nu }\partial _{\nu }\xi )=u^{\nu }\partial _{\mu }\partial _{\nu }\xi
+\left( \partial _{\mu }u^{\nu }\right) \partial _{\nu }\xi  \nonumber \\
&=&u^{\nu }\partial _{\nu }\partial _{\mu }\xi -\frac{\varepsilon +P}{n}%
u_{\nu }\left( \partial _{\mu }u^{\nu }\right)  \nonumber \\
&=&\frac{\partial }{\partial \tau }\partial _{\mu }\xi -\frac{\varepsilon +P%
}{n}\partial _{\mu }\left( \frac{1}{2}u_{\nu }u^{\nu }\right) =\frac{%
\partial }{\partial \tau }\partial _{\mu }\xi ,
\end{eqnarray}
that is, the two derivatives $\partial _{\mu }$ and $\partial /\partial \tau 
$ commute when applied to $\xi $. Therefore, from Eqs.(\ref{divl}) and (\ref
{dldtau}) we can eliminate $\xi $ to get 
\begin{equation}
u^{\nu }\partial _{\nu }\left[ \frac{\varepsilon +P}{n}u_{\mu }\right] =%
\frac{\partial _{\mu }P}{n}.
\end{equation}
Again using Eq.(\ref{adiab2}), this reduces to 
\begin{equation}
u_{\mu }\,u^{\nu }(\partial _{\nu }P)+\left( \varepsilon +P\right) u^{\nu
}\partial _{\nu }\,u_{\mu }=\partial _{\mu }P.  \label{ff}
\end{equation}
The first term of this equation is further modified as 
\begin{eqnarray}
u^{\nu }(\partial _{\nu }P) &=&u^{\nu }n\partial _{\nu }\left[ \frac{%
\varepsilon +P}{n}\right]  \nonumber \\
&=&u^{\nu }\partial _{\nu }\left( \varepsilon +P\right) -\frac{\varepsilon +P%
}{n}u^{\nu }\partial _{\nu }n  \nonumber \\
&=&u^{\nu }\partial _{\nu }\left( \varepsilon +P\right) +\left( \varepsilon
+P\right) \partial _{\nu }u^{\nu },
\end{eqnarray}
where the continuity equation (\ref{conti}) was used. Thus, Eq.(\ref{ff})
can be rewritten as 
\begin{equation}
\partial _{\nu }\,T_{\;\mu }^{\nu }=0,  \label{Conserv}
\end{equation}
where 
\begin{equation}
T_{\mu \nu }=\left( \varepsilon +P\right) u_{\mu }u_{\nu }-Pg_{\mu \nu }\;.
\end{equation}
That is, we arrive at the equation of motion of relativistic fluid dynamics
with the energy-momentum tensor of the perfect fluid. From this equation, we
obtain the relativistic version of the Euler equation\cite{Weinberg}, 
\begin{equation}
\frac{\partial }{\partial t}\vec{v}+\left( \vec{v}\cdot \nabla \right) \,%
\vec{v}=-\frac{1}{\left( \varepsilon +P\right) \gamma ^{2}}\left[ \nabla P+%
\vec{v}\frac{\partial P}{\partial t}\right] .
\end{equation}

In the above derivation, we assumed the Minkowski space-time metric, but as
mentioned before all the calculations can easily be extended to the case
where the metric is more general. For example, the final result Eq.(\ref
{Conserv}) in the curved metric is 
\begin{equation}
\,T_{\;\mu \;;\nu }^{\nu }=0,  \label{Conserv_g}
\end{equation}
where $;$ stands for the covariant derivative, as usual.

As pointed out by several authors\cite{Schutz,Ray,Chiueh}, the above scheme
leads only to non-rotational flow. This can be seen from Eq.(\ref{c}), where
the velocity field is proportional to the four-gradient of a scalar
function. In order to include the rotational flow, we have to add a term
coming from another constraint with respect to entropy in the original
action. However, for the spherically symmetric case below, one does not need
to worry about the rotational flow, so we omit the discussion for the sake
of simplicity.

\section{Spherically Symmetric Case}

The above variational approach is particularly useful when we can solve the
constraint equations explicitly. One dimensional, or spherically symmetric
system is such a case. Here we study the spherically symmetric case. Let the
density profile in a space fixed frame be 
\begin{equation}
\bar{n}=f(r,t).  \label{f}
\end{equation}
Then the velocity field is determined from the continuity equation as 
\begin{equation}
v=-\frac{1}{r^{2}f}\int_{0}^{r}r^{2}\dot{f}dr.  \label{vprofil}
\end{equation}
If we use this expression for the velocity field, then we can omit the
constraint terms in the action. Thus, we have a Lagrangian for $f$ as 
\begin{equation}
L=-4\pi \int_{0}^{\infty }r^{2}dr\;\varepsilon \left( n\right) ,
\label{Lagrange}
\end{equation}
where the local density in the comoving frame is given by 
\begin{equation}
n=\frac{f}{\gamma }.  \label{dens}
\end{equation}
To see explicitly how the variational principle works with this Lagrangian,
let us consider an arbitrary density variation, 
\[
f\rightarrow f+\delta f.
\]
Under such variation, we get 
\begin{eqnarray}
\delta I &=&-4\pi \int dt\int_{0}^{\infty }r^{2}dr\;\delta \varepsilon
\left( \frac{1}{\gamma }f\right)   \nonumber \\
&=&4\pi \int dt\int_{0}^{\infty }r^{2}dr\;\left( \frac{\delta \varepsilon }{%
\delta n}\right) \left( -\frac{\delta f\,}{\gamma }+v\gamma f\delta v\right)
,
\end{eqnarray}
and, 
\[
\delta v=-\frac{\delta f}{f}v-\frac{1}{r^{2}f}\int_{0}^{r}r^{2}\delta \dot{f}%
dr,
\]
\begin{eqnarray}
\delta I &=&4\pi \int dt\int_{0}^{\infty }r^{2}dr\;\left( \frac{\delta
\varepsilon }{\delta n}\right) \left( -\frac{\delta f\,}{\gamma }+v\gamma
\left\{ -\delta f\,v-\frac{1}{r^{2}}\int_{0}^{r}r^{\prime \,2}\delta \dot{f}%
dr^{\prime }\right\} \right)   \nonumber \\
&=&-4\pi \int dt\left[ \int_{0}^{\infty }r^{2}dr\;\left( \frac{\delta
\varepsilon }{\delta n}\right) \gamma \delta f+\int_{0}^{\infty }dr\;\left( 
\frac{\delta \varepsilon }{\delta n}\right) v\gamma \int_{0}^{r}r^{\prime
\,2}\delta \dot{f}dr^{\prime }\right]   \nonumber \\
&=&-4\pi \int dt\int_{0}^{\infty }r^{2}dr\left\{ \left( \frac{\delta
\varepsilon }{\delta n}\right) \gamma -\frac{\partial }{\partial t}\left[
\int_{r}^{\infty }dr\left( \frac{\delta \varepsilon }{\delta n}\right)
v\gamma \right] \right\} \delta f
\end{eqnarray}
From this we have 
\[
\left( \frac{\delta \varepsilon }{\delta n}\right) \gamma -\frac{\partial }{%
\partial t}\left[ \int_{r}^{\infty }dr\left( \frac{\delta \varepsilon }{%
\delta n}\right) v\gamma \right] =0.
\]
Taking the derivative with respect to $r$ of the both sides, we get 
\[
\frac{\partial }{\partial t}\left[ \left( \frac{\delta \varepsilon }{\delta n%
}\right) v\gamma \right] =-\frac{\partial }{\partial r}\left[ \left( \frac{%
\delta \varepsilon }{\delta n}\right) \gamma \right] .
\]
Using the adiabatic relation of the energy density and pressure in the
equation of motion, we get 
\begin{equation}
\dot{v}+v\frac{\partial v}{\partial r}=-\frac{1}{(\varepsilon +P)\gamma ^{2}}%
\left\{ \frac{\partial P}{\partial r}+v\frac{\partial P}{\partial t}\right\} 
\label{eqm1}
\end{equation}
which is again the relativistic Euler equation for spherically symmetric case
\cite{Weinberg}.

\subsection{Effective Lagrangian}

Now let us introduce a parametric ansatz for the density profile as 
\begin{equation}
\bar{n}(t,r)=f(r,a(t)),  \label{nparam}
\end{equation}
where $a=a(t)$ is a (set of) time-dependent parameter(s) which determines
the form of $f$. We suppose that $f$ is suitably normalized, 
\begin{equation}
4\pi \int_{0}^{\infty }r^{2}fdr=N.
\end{equation}
The velocity profile, Eq.(\ref{vprofil}) becomes 
\begin{equation}
v(r,t)=-\dot{a}\frac{1}{r^{2}f}\int_{0}^{r}r^{2}\left( \frac{\partial f}{%
\partial a}\right) dr\equiv \Delta \;\dot{a},  \label{vparam}
\end{equation}
with 
\begin{equation}
\Delta =\Delta (r,a)=-\frac{1}{r^{2}f}\int_{0}^{r}r^{\prime \,2}\left( \frac{%
\partial f}{\partial a}\right) dr^{\prime }.  \label{Delta}
\end{equation}
For more than one parameter, $a=\left\{ a^{i},i=1,...,n\right\} $ Eqs.(\ref
{vprofil}) and (\ref{Delta}) should be understood as 
\begin{equation}
v(r,t)=\sum_{i=1}^{n}\Delta _{i}\dot{a}^{i},\;\Delta _{i}=-\frac{1}{r^{2}f}%
\int_{0}^{r}r^{\prime \,2}\left( \frac{\partial f}{\partial a^{i}}\right)
dr^{\prime }.
\end{equation}
The important point here is that the velocity field is linear in $\dot{a}$.
The effective Lagrangian for our variable $a=a(t)$ becomes 
\begin{equation}
L(a,\dot{a})=-4\pi \int_{0}^{\infty }r^{2}dr\;\varepsilon \left( n\right) ,
\label{Leff-1}
\end{equation}
where $n=f(r,a)/\gamma $ and $\gamma =1/\sqrt{1-\left( \Delta \;\dot{a}%
\right) ^{2}}$. The equation of motion for the variable $a$ is obtained from
the Euler-Lagrange equation 
\begin{equation}
\frac{d}{dt}\left( \frac{\partial L}{\partial \dot{a}}\right) -\frac{%
\partial L}{\partial a}=0,
\end{equation}
which is written as 
\begin{equation}
4\pi \int_{0}^{\infty }r^{2}dr\left[ \frac{d}{dt}\left\{ \frac{\delta
\varepsilon }{\delta n}n\gamma ^{2}\Delta ^{2}\dot{a}\right\} +\frac{\delta
\varepsilon }{\delta n}\frac{\partial n}{\partial a}\right] =0,
\end{equation}
For adiabatic motion, we get 
\begin{equation}
\frac{d}{dt}\int_{0}^{\infty }r^{2}dr\left\{ \left( \varepsilon +P\right)
\gamma ^{2}v\Delta \right\} =-\int_{0}^{\infty }r^{2}dr\left( \varepsilon
+P\right) \left( \frac{1}{f}\frac{\partial f}{\partial a}-\gamma ^{2}v\frac{%
\partial v}{\partial a}\right) .  \label{euler-1}
\end{equation}
The effective Hamiltonian $H$ is then 
\begin{eqnarray}
H &\equiv &\dot{a}\frac{\partial L}{\partial \dot{a}}-L  \nonumber \\
&=&4\pi \int r^{2}dr\;\left[ -\frac{\delta \varepsilon }{\delta n}\bar{n}%
\dot{a}\frac{\partial }{\partial \dot{a}}\left( \frac{1}{\gamma }\right)
+\varepsilon \right] =4\pi \int r^{2}dr\;\left[ \left( \varepsilon +P\right)
\gamma ^{2}v^{2}+\varepsilon \right]  \nonumber \\
&=&4\pi \int r^{2}dr\;\left[ \left( \varepsilon +P\right) \gamma ^{2}-P%
\right] =\int d^{3}r\;T^{00},  \label{H}
\end{eqnarray}
which is in fact the total energy of the system and a conserved quantity.

\subsection{Relativistic Rayleigh-Plesset Equation}

For an example, let us consider a system composed of a homogeneous spherical
gas bubble surrounded by a homogeneous fluid. We then introduce the ansatz, 
\begin{eqnarray}
f &=&\rho _{0G}=\frac{3}{4\pi }\frac{N_{G}}{R^{3}}, 
\begin{array}{ll}
& 0<r<R
\end{array}
\nonumber \\
&=&\rho _{0L}=\frac{3}{4\pi }\frac{N_{L}}{R_{\infty }^{3}-R^{3}}, 
\begin{array}{ll}
& R<r<R_{\infty }
\end{array}
\end{eqnarray}
where the radius of the gas bubble $R=R(t)$ is the only dynamic variable. We
use the subscript $G$ and $L$ to specify the quantities in the gas and the
fluid, respectively. For example, $N_{G}$ and $N_{L}$ are number of
particles (constant) in the gas and fluid, respectively. The outer
(constant) radius of the fluid $R_{\infty }$ is introduced here to take into
account the conservation of the number of particles in the fluid, but
ultimately should be taken equal to $\infty $. The velocity field is then
determined as 
\begin{eqnarray}
v(r) &=&\frac{r}{R}\dot{R}, 
\begin{array}{ll}
\;\;\;\;\;\;\;\;\;\;\;\; & 0<r<R,
\end{array}
\nonumber \\
&=&\frac{R^{2}\left( R_{\infty }^{3}-r^{3}\right) }{r^{2}\left( R_{\infty
}^{3}-R^{3}\right) }\dot{R}, 
\begin{array}{ll}
& R<r<R_{\infty },
\end{array}
\end{eqnarray}
so that 
\begin{eqnarray}
\Delta (r) &=&\frac{r}{R}, 
\begin{array}{ll}
\;\;\;\;\;\;\;\;\;\;\;\; & 0<r<R,
\end{array}
\nonumber \\
&=&\frac{R^{2}\left( R_{\infty }^{3}-r^{3}\right) }{r^{2}\left( R_{\infty
}^{3}-R^{3}\right) }, 
\begin{array}{ll}
& R<r<R_{\infty }.
\end{array}
\end{eqnarray}
The effective Lagrangian for $R$ is then 
\begin{equation}
L=-4\pi \int_{0}^{R}r^{2}dr\;\varepsilon _{G}\left( f_{G}\sqrt{1-\left( 
\frac{r}{R}\dot{R}\right) ^{2}}\right) -4\pi \int_{R}^{R_{\infty
}}r^{2}dr\;\varepsilon _{L}\left( f_{L}\sqrt{1-\left( \frac{R^{2}\left(
R_{\infty }^{3}-r^{3}\right) }{r^{2}\left( R_{\infty }^{3}-R^{3}\right) }%
\dot{R}\right) ^{2}}\right) .  \label{L-RP}
\end{equation}
The equation of motion for $R=R(t)$ is given by 
\begin{equation}
\frac{d}{dt}\left( \frac{\partial L}{\partial \dot{R}}\right) =\frac{%
\partial L}{\partial R},
\end{equation}
where 
\begin{equation}
\frac{\partial L}{\partial \dot{R}}=4\pi \frac{\dot{R}}{R^{2}}%
\int_{0}^{R}dr\,r^{4}\left( \varepsilon _{G}+P_{G}\right) \gamma ^{2}+4\pi
R^{4}\dot{R}\int_{R}^{R_{\infty }}\frac{dr}{r^{2}}\;\left( \varepsilon
_{L}+P_{L}\right) \gamma ^{2}\left( \frac{R_{\infty }^{3}-r^{3}}{R_{\infty
}^{3}-R^{3}}\right) ^{2},  \label{dldrdot}
\end{equation}
and 
\begin{eqnarray}
\frac{\partial L}{\partial R} &=&-4\pi R^{2}\left[ \varepsilon
_{G}-\varepsilon _{L}\right] _{R}-4\pi \int_{0}^{R}r^{2}dr\;\left(
\varepsilon _{G}+P_{G}\right) \left( -\frac{3}{R}+\gamma ^{2}\left( \frac{%
\dot{R}}{R}\right) ^{2}\frac{r^{2}}{R}\right)  \nonumber \\
&&-4\pi \int_{R}^{R_{\infty }}r^{2}dr\left( \varepsilon _{L}+P_{L}\right)
\left( \frac{3R^{2}}{R_{\infty }^{3}-R^{3}}-\gamma ^{2}\frac{R^{3}\left(
2R_{\infty }^{3}+R^{3}\right) \left( R_{\infty }^{3}-r^{3}\right) ^{2}}{%
r^{4}\left( R_{\infty }^{3}-R^{3}\right) ^{3}}\dot{R}^{2}\right) .  \nonumber
\\
&&  \label{dldr}
\end{eqnarray}
At this stage, we can take the limit $R_{\infty }\rightarrow \infty $. Thus,
we have 
\begin{equation}
\frac{\partial L}{\partial \dot{R}}=4\pi \frac{\dot{R}}{R^{2}}%
\int_{0}^{R}dr\,r^{4}\left( \varepsilon _{G}+P_{G}\right) \gamma ^{2}+4\pi
R^{4}\dot{R}\int_{R}^{\infty }\frac{dr}{r^{2}}\;\left( \varepsilon
_{L}+P_{L}\right) \gamma ^{2},
\end{equation}
and 
\begin{eqnarray}
\frac{\partial L}{\partial R} &=&-4\pi R^{2}\left( \left[ \varepsilon
_{G}-\varepsilon _{L}\right] _{R}+\left[ \varepsilon _{L}+P_{L}\right]
_{\infty }\right) +\frac{12\pi }{R}\int_{0}^{R}r^{2}dr\;\left( \varepsilon
_{G}+P_{G}\right)  \nonumber \\
&&-4\pi \int_{0}^{R}r^{2}dr\;\left( \varepsilon _{G}+P_{G}\right) \gamma
^{2}\left( \frac{\dot{R}}{R}\right) ^{2}\frac{r^{2}}{R}+8\pi R^{3}\dot{R}%
^{2}\int_{R}^{\infty }\frac{dr}{r^{2}}\left( \varepsilon _{L}+P_{L}\right)
\gamma ^{2}.  \nonumber \\
&&
\end{eqnarray}
The equation of motion then takes the form, 
\begin{equation}
\frac{d}{dt}\left[ \left( I_{1}+I_{2}\right) R^{3}\dot{R}\right] =F-\left(
I_{1}-2I_{2}\right) R^{2}\dot{R}^{2},  \label{RP-rel}
\end{equation}
where 
\begin{eqnarray}
I_{1} &=&\int_{0}^{1}dx\;x^{4}\left( \varepsilon _{G}+P_{G}\right) \gamma
^{2},  \label{I1} \\
I_{2} &=&\int_{1}^{\infty }\frac{dx}{x^{2}}\left( \varepsilon
_{L}+P_{L}\right) \gamma ^{2},  \label{I2}
\end{eqnarray}
with $x=r/R$ and 
\begin{equation}
F=-R^{2}\left( \left[ \varepsilon _{G}-\varepsilon _{L}\right] _{R}+\left[
\varepsilon _{L}+P_{L}\right] _{0}\right) +\frac{3}{R}\int_{0}^{R}r^{2}dr\;%
\left( \varepsilon _{G}+P_{G}\right) ,  \label{DP}
\end{equation}
where the subscript $0$ represents the quantity evaluated at $v=0$. Eq.(\ref
{RP-rel}) is a full relativistic equation of motion for the radius of a gas
bubble under the homologous motion of the system.

To see the non-relativistic limit of Eq.(\ref{RP-rel}), we separate the
energy density into the sum of the rest-mass energy density $\rho $ and the
internal energy density $\varepsilon ^{int}$ as 
\begin{equation}
\varepsilon =\rho +\varepsilon ^{int},
\end{equation}
and expand Eqs.(\ref{I1},\ref{I2},\ref{DP}) in a power series of small
parameters such as , $v^{2},\varepsilon _{int}/\rho ,$and $P/\rho $ in the
non-relativistic regime. We have 
\begin{equation}
\left( \varepsilon +P\right) \gamma ^{2}\simeq \rho _{0}+\varepsilon
_{0}^{int}+P_{0}+\frac{1}{2}\rho _{0}v^{2},
\end{equation}
Thus, 
\begin{eqnarray}
I_{1} &\simeq &\frac{1}{5}\left( \rho _{0,G}+\varepsilon
_{0,G}^{int}+P_{0,G}\right) +\frac{1}{14}\rho _{0,G}\dot{R}^{2}, \\
I_{2} &\simeq &\rho _{0,L}+\varepsilon _{0,L}^{int}+P_{0,L}+\frac{1}{10}\rho
_{0,L}\dot{R}^{2},
\end{eqnarray}
and 
\begin{eqnarray}
F &\simeq &R^{2}\left( P_{0,G}-P_{0,L}\right) -\left( \frac{1}{2}\left( \rho
_{0,L}+P_{0,L}\right) +\frac{3}{10}\left( \rho _{0,G}+P_{0,G}+c_{s}^{2}\rho
_{0,G}\right) \right) R^{2}\dot{R}^{2}  \nonumber \\
&&-\frac{1}{8}\left( \rho _{0,L}+\frac{3}{7}\rho _{G,0}\right) R^{2}\dot{R}%
^{4}.
\end{eqnarray}
In the lowest order in $\dot{R}$ we get 
\begin{equation}
\left( \frac{1}{5}\rho _{0,G}+\rho _{0,L}\right) R\ddot{R}+\frac{3}{2}\rho
_{0,L}\dot{R}^{2}=P_{0,G}-P_{0,L},  \label{RP-nr}
\end{equation}
which is the usual Rayleigh-Plesset equation of a gas bubble inside a liquid
\cite{KTR} without the term for the energy dissipation due to the sound
radiation.

When the equation of state of the fluid and the gas are given as 
\begin{equation}
P\propto \rho ^{\Gamma }
\end{equation}
where $\Gamma $ is the adiabatic index, all the integrals in Eq.(\ref{RP-rel}%
) are expressed analytically. We get 
\begin{eqnarray}
&&MR^{3}\ddot{R}+\left\{ \frac{1}{5}\rho _{0,G}J_{1}+\rho _{0,L}J_{3}+\frac{%
4-3\Gamma _{G}}{5}\left( \varepsilon ^{in}+P\right) _{0,G}J_{2}+\left(
\varepsilon ^{in}+P\right) _{0,L}J_{4}\right\} R^{2}\dot{R}^{2}  \nonumber \\
&=&-R^{2}\left( \left[ \varepsilon _{G}-\varepsilon _{L}\right] _{R}+\left[
\varepsilon _{L}+P_{L}\right] _{0}\right) +R^{2}\left( \rho
_{0,G}J_{9}+\left( \varepsilon ^{in}+P\right) _{0,G}J_{10}\right) ,
\label{RP-gm}
\end{eqnarray}
with 
\begin{eqnarray}
M &=&\frac{1}{5}\rho _{0,G}J_{1}+\rho _{0,L}J_{3}+\frac{1}{5}\left(
\varepsilon ^{in}+P\right) _{0,G}J_{2}+\left( \varepsilon ^{in}+P\right)
_{0,L}J_{4}  \nonumber \\
&&+\left[ \frac{1}{7}\left( \rho _{0,G}J_{5}+\left( 2-\Gamma \right) \left(
\varepsilon ^{in}+P\right) _{0,G}J_{6}\right) +\frac{1}{5}\left( \rho
_{0,L}J_{7}+\left( 2-\Gamma \right) \left( \varepsilon ^{in}+P\right)
_{0,L}J_{8}\right) \right] \dot{R}^{2}  \nonumber \\
&&
\end{eqnarray}
and 
\begin{eqnarray*}
J_{1} &=&F\left( \left[ \frac{1}{2},\frac{5}{2}\right] ,\frac{7}{2},\dot{R}%
^{2}\right) ,\;J_{2}=F\left( \left[ 1-\frac{\Gamma }{2},\frac{5}{2}\right] ,%
\frac{7}{2},\dot{R}^{2}\right) , \\
J_{3} &=&F\left( \left[ \frac{1}{2},\frac{1}{4}\right] ,\frac{5}{4},\dot{R}%
^{2}\right) ,\;J_{4}=F\left( \left[ 1-\frac{\Gamma }{2},\frac{1}{4}\right] ,%
\frac{5}{4},\dot{R}^{2}\right) , \\
J_{5} &=&F\left( \left[ \frac{3}{2},\frac{7}{2}\right] ,\frac{9}{2}%
,R^{2}\right) ,\;J_{6}=F\left( \left[ 2-\frac{\Gamma }{2},\frac{7}{2}\right]
,\frac{9}{2},\dot{R}^{2}\right) , \\
J_{7} &=&F\left( \left[ \frac{3}{2},\frac{5}{4}\right] ,\frac{9}{4},\dot{R}%
^{2}\right) ,\;J_{8}=F\left( \left[ 2-\frac{\Gamma }{2},\frac{5}{4}\right] ,%
\frac{9}{4},\dot{R}^{2}\right) , \\
J_{9} &=&F\left( \left[ \frac{3}{2},-\frac{1}{2}\right] ,\frac{5}{2},\dot{R}%
^{2}\right) ,J_{10}=F\left( \left[ \frac{3}{2},-\frac{\Gamma }{2}\right] ,%
\frac{5}{2},\dot{R}^{2}\right) ,
\end{eqnarray*}
where $F([a,b],c,z)$ is the hypergeometric function.

For the sake of illustration, we show in Figs.1, 2, and 3, time dependences
of the radius and velocity described by the relativistic Rayleigh-Plesset
equation Eq.(\ref{RP-gm}) for 3 different initial conditions. In this
example, we consider the case where the both gas and fluid have the same
mass density and the adiabatic index $\Gamma =4/3$. Three cases shown here
are for the different values of the initial gas pressure, $\left( P/\rho
\right) _{G,0}=1/10$ (Fig. 1)$,5$ (Fig. 2)$,$ and $100$ (Fig. 3), keeping
the ratio of the initial gas to the liquid pressure $P_{G,0}/P_{L,0}=100$.
Solid lines are for the full relativistic equation of motion (Eq.\ref{RP-gm}%
) and the dashed ones are for the non-relativistic limit, Eq.(\ref{RP-nr}).
For the low initial gas pressure, two solutions coincide (Fig. 1). For the
extremely high initial pressure, the motion of the bubble becomes completely
relativistic (Fig. 3) and the non-relativistic equation of motion differs
completely from the relativistic equation. Note that in the relativistic
equation, the velocity saturates at $v/c=1$. Of course, in this extreme
example, the fluid motion becomes supersonic $\left( v>v_{s}=\sqrt{1/3}%
=0.577\right) $ and the hypothesis of homologous motion may breakdown.
However, it is important to note that, there exists a case where the
non-relativistic approximation fails down completely although the fluid and
gas motion are still subsonic like in the case of Fig.2.

\vspace{0.5cm} 
\begin{figure}[htbp]
\end{figure}

\section{Nonadiabatic processes}

In many cases, the change of the density associated with the flow of the
matter causes non-quasi static processes within the hydrodynamical volume
element established in the practical calculations. For example, in the limit
of large Reynolds number, the dynamical change of the volume easily leads to
a highly turbulent regime in small regions of the fluid, and this complex
fluid motion will gradually thermalize inside the volume element. In such a
case, there appears the heat production inside of such a volume element. If
the time scale for the thermalization is negligible, then the heat
production can be expressed in terms of viscous tensor, and the hydrodynamic
equation of motion becomes, 
\begin{equation}
\partial _{\mu }\left( T^{\mu \nu }+\Sigma ^{\mu \nu }\right) =0,
\label{stress}
\end{equation}
where $\Sigma ^{\mu \nu }$ is the shear tensor. When there is no heat
transfer, we may take 
\begin{equation}
\Sigma ^{\mu \nu }=q(u^{\mu }u^{\nu }-g^{\mu \nu }),  \label{Sigma}
\end{equation}
where $q$ is a function of the local thermodynamical quantities, like $\rho
,P$ and its derivatives. From Eq.(\ref{stress}) we have 
\begin{equation}
u^{\mu }\partial _{\mu }\left( \frac{\varepsilon }{n}\right) +(P+q)u^{\mu
}\partial _{\mu }\left( \frac{1}{n}\right) =0.
\end{equation}
This means that, in the Lagrange comoving system, the change of the specific
energy with respect to the proper time is given by 
\begin{equation}
\frac{dE}{d\tau }+P\frac{dV}{d\tau }=-q\frac{dV}{d\tau }.  \label{dedt}
\end{equation}
Thus we identify the function $q$ as the rate of the production of the
entropy $S$ with respect to the volume change, 
\begin{equation}
q=-T\frac{dS}{d\tau }/\frac{dV}{d\tau }  \label{dsdt}
\end{equation}
The specific form of $\Sigma ^{\mu \nu }$, Eq.(\ref{Sigma}) allows us to
write 
\begin{equation}
\tilde{T}^{\mu \nu }=\left( \varepsilon +P+q\right) u^{\mu }u^{\nu }-\left(
P+q\right) g^{\mu \nu },
\end{equation}
which conserves 
\begin{equation}
\partial _{\mu }\tilde{T}^{\mu \nu }=0.
\end{equation}
From the above conservation law we get immediately the equation of motion, 
\begin{equation}
\dot{v}+v\frac{\partial v}{\partial r}=-\frac{1}{(\varepsilon +P+q)\gamma
^{2}}\left\{ \frac{\partial (P+q)}{\partial r}+v\frac{\partial (P+q)}{%
\partial t}\right\} ,  \label{eqm2}
\end{equation}
which describes the relativistic hydrodynamical motion under the local
entropy production, Eq.(\ref{dedt}). The function $q$ should be specified
appropriately according to the non-adiabatic processes representing the
conversion of kinetic energy of the collective motion to the internal energy
of the matter. Such a viscosity was first introduced in the non-relativistic
hydrodynamics by Neumann and Richtmyer\cite{Neuman} in order to simulate the
entropy production mechanism at the shock front. Eq.(\ref{eqm2}) is the
relativistic extension of the method of pseudo-viscosity of Neumann and
Richtmyer.

The above scheme is easily incorporated in the variational formalism. In the
presence of non-adiabatic processes, the variation in the specific energy in
the previous section, Eq.(\ref{dE=pdV}), should be replaced by 
\begin{equation}
\delta E=-P\delta V+\delta Q=-P\delta V+T\delta S=-\left( P+q\right) \delta V
\label{thermo2}
\end{equation}
where $\delta Q$ is the generated heat associated with the non quasi-static
density variation. Consequently we should, instead of the adiabatic
relations (\ref{adiab1},\ref{adiab2}), use 
\begin{eqnarray}
\frac{\partial \varepsilon }{\partial n} &\rightarrow &\frac{\varepsilon +P+q%
}{n},  \label{adiab3} \\
d\left( \frac{\delta \varepsilon }{\delta n}\right) &\rightarrow &\frac{1}{n}%
d(P+q).  \label{adiab4}
\end{eqnarray}
From these substitutions, we get immediately Eq.(\ref{eqm2}). In terms of
parametric representation, the equation of motion is given by 
\begin{equation}
\frac{d}{dt}\int_{0}^{\infty }r^{2}dr\left\{ \left( \varepsilon +P+q\right)
\gamma ^{2}v\Delta \right\} =-\int_{0}^{\infty }r^{2}dr\left( \varepsilon
+P+q\right) \left( \frac{1}{f}\frac{\partial f}{\partial a}-\gamma ^{2}v%
\frac{\partial v}{\partial a}\right) .  \label{eqm3}
\end{equation}
The effective Hamiltonian for dynamical variable $a$ is given again by 
\begin{eqnarray}
H &\equiv &\dot{a}\frac{\partial L}{\partial \dot{a}}-L  \nonumber \\
&=&4\pi \int r^{2}dr\;\left[ \left( \varepsilon +P+q\right) \gamma
^{2}-\left( P+q\right) \right] =\int d^{3}r\;\left( T^{00}+\Sigma
^{00}\right) ,
\end{eqnarray}
which is conserved, 
\begin{equation}
\frac{dH}{dt}=0,
\end{equation}
for the equation of motion, Eq.(\ref{eqm3}), together with Eq.(\ref{dedt}).
This $H$ can again be identified as the total energy of the system including
the internal heat energy generated in the fluid.

\section{General Relativistic Hydrodynamics}

For the application of the present formalism to astrophysical problems it is
essential to include the effect of gravity through the theory of General
Relativity. The variational approach of the general relativistic
hydrodynamics has been discussed by several authors\cite
{Taub,Schutz,Ray,Bogos}. In this section, starting from the variational
approach, we show that the method of effective Lagrangian can also be
established taking the metric as one of the variational trial functions. Let
us first review how the general relativistic energy and momentum tensor are
derived from the variational approach.

\subsection{Energy-Momentum Tensor and Einstein's Equation}

The total action is given as 
\[
I=I_{G}+I_{M}, 
\]
where 
\begin{equation}
I_{G}=\frac{1}{2\kappa }\int d^{4}x\sqrt{-g}{\cal \Re },
\end{equation}
is the action for the gravitational field and $\kappa =8\pi G$ with $G$ the
gravitational constant. As usual, $g=\det \left| g_{\mu \nu }\right| ,$ is
the determinant of the metric tensor $g_{\mu \nu }$, and ${\cal \Re }$ is
the curvature scalar. The action of matter is now given by\cite{Taub} 
\begin{equation}
I_{M}=\int d^{4}x\sqrt{-g}\left\{ -\varepsilon (n)+\xi (x)\left( nu^{\mu
}\right) _{;\mu }+\;\frac{1}{2}\zeta (x)\left( u^{\mu }u_{\mu }-1\right)
\right\} ,  \label{I}
\end{equation}
where $\xi $ and $\zeta /2$ are Lagrange multipliers as before. As usual, ``$%
;$'' represents the covariant derivative and the factor $\sqrt{-g}$ is
inserted to guarantee that the Lagrangian density is a scalar. The variation
of the action should be carried out with respect to $g_{\mu \nu },n,u^{\mu
},\xi ,\,$and $\zeta ,$ independently. The results of variations with
respect to $n,u^{\mu },\xi ,$ and $\zeta $ are the same as before (see Eqs.(%
\ref{a}),(\ref{c}),(\ref{zeta}), (\ref{divl}) , (\ref{Conserv_g}) and the
comments for the covariant derivative in Sec.II). Thus, these variations
gives the relativistic hydrodynamic equation for a given metric $g^{\mu \nu
} $.

The functional derivative with respect to $g_{\mu \nu }$ is calculated to be 
\begin{equation}
\frac{\delta I_{M}}{\delta g_{\mu \nu }}=-\frac{\partial \sqrt{-g}}{\partial
g_{\mu \nu }}\left\{ \varepsilon (n)+n\left( \partial _{\mu }\xi (x)\right)
u^{\mu }\right\} +\frac{1}{2}\zeta \sqrt{-g}u^{\mu }u^{\nu },
\end{equation}
where we already employed the constraints, cf. Eqs.(\ref{uu=1},\ref{conti}).
Using 
\begin{equation}
\frac{\partial \sqrt{-g}}{\partial g_{\mu \nu }}=-\frac{1}{2}\sqrt{-g}g^{\mu
\nu },
\end{equation}
and substituting the values of $\partial _{\mu }\xi $ and $\zeta $, we get, 
\begin{eqnarray}
\frac{\delta I_{M}}{\delta g_{\mu \nu }} &=&\frac{1}{2}\sqrt{-g}g^{\mu \nu
}\left\{ \varepsilon (n)-n\left( \frac{\varepsilon +P}{n}u_{\mu }\right)
u^{\mu }\right\} +\frac{1}{2}\sqrt{-g}\left( \varepsilon +P\right) u^{\mu
}u^{\nu }  \nonumber \\
&=&\frac{1}{2}\sqrt{-g}\left\{ \left( \varepsilon +P\right) u^{\mu }u^{\nu
}-Pg^{\mu \nu }\right\} .
\end{eqnarray}
Comparing this result to the definition of the energy-momentum tensor, 
\begin{equation}
\frac{\delta I_{M}}{\delta g_{\mu \nu }}\equiv \frac{1}{2}\sqrt{-g}T^{\mu
\nu },
\end{equation}
we identify that 
\begin{equation}
T^{\mu \nu }=\left( \varepsilon +P\right) u^{\mu }u^{\nu }-Pg^{\mu \nu },
\end{equation}
which is nothing but the energy-momentum tensor of the perfect fluid. Thus,
the energy-momentum tensor of the fluid is derived from the Lagrangian
density Eq.(\ref{I}) just as in the case of the field theoretical
Lagrangian. Note that the role of constraints are essential for this
derivation.

The variation of the gravitational action $I_{G}$ with respect to $g_{\mu
\nu }$ gives the usual Einstein tensor ${\cal G}^{\mu \nu }$, 
\[
\frac{\delta I_{G}}{\delta g_{\mu \nu }}\equiv -\frac{1}{2}\sqrt{-g}{\cal G}%
^{\mu \nu }, 
\]
so that we get 
\begin{equation}
{\cal G}^{\mu \nu }=\kappa T^{\mu \nu },  \label{Einstein}
\end{equation}
which is the Einstein equation, as expected. The hydrodynamical equation,
Eq.(\ref{Conserv_g}), 
\begin{equation}
T_{\;\;\;;\mu }^{\mu \nu }=0,
\end{equation}
can be re-obtained from Eq.(\ref{Einstein}) due to the Bianchi identity, 
\begin{equation}
{\cal G}_{\;\;\;;\mu }^{\mu \nu }=0.  \label{Bian}
\end{equation}
It is interesting to note that if we use the metric functions as basic
dynamical variables then, the hydrodynamic equation of motion is obtained
somewhat indirectly from the properties of metric tensor and the variational
principle does not lead directly to the equation of motion. This point will
be discussed later again in the context of the derivation of the special
relativistic equation of motion using the comoving coordinate system.

\subsection{Spherically Symmetric System}

The derivation of the equation of motion above is too formal and not much
useful to be applied directly for some practical problems. To make use of
the variational approach, it is necessary to establish appropriate trial
functions in order to write down the effective Lagrangian for these
functions. As in the case of special relativity, this is possible when the
system has appropriate symmetry, such as spherically symmetric distribution
of matter. Many problems of the gravitational collapse of stars, the
structure of neutron stars, and the Robertson-Walker cosmology can be
discussed in this symmetry. Here we establish the effective Lagrangian for
the spherically symmetric system.

The most general form of the metric for a spherically symmetric system can
be taken as\cite{Weinberg} 
\begin{equation}
ds^{2}=e^{2\phi }dT^{2}-e^{2\lambda }d\xi ^{2}-r^{2}d\Omega ^{2},
\label{metric}
\end{equation}
where $\left( T,\xi \right) $ denotes the time and radial coordinates and $%
\phi =\phi (\xi ,T),\;\lambda =\lambda (\xi ,T),\;$and$\;r=r(\xi ,T)$ are
unknown functions to be determined. Usually, if we consider the radial
velocity field of the fluid as an independent variable, then we need only
two independent functions in the metric and we may choose, for example, $%
r=\xi $. However, with the above metric involving three functions, we can
further take the so-called comoving frame in such a way that the space-like
components of the four-velocity field vanish everywhere\cite{Ellis}, 
\begin{equation}
u^{\mu }=(u^{0},0,0,0).  \label{u}
\end{equation}
From the normalization condition $u_{\mu }u^{\mu }=1$, we get 
\begin{equation}
u^{0}=e^{-\phi }.  \label{u0}
\end{equation}
In this comoving frame, the conservation law is expressed as 
\begin{equation}
\left( nu^{\mu }\right) _{;\mu }=\frac{1}{\sqrt{-g}}\partial _{\mu }\left( 
\sqrt{-g}nu^{\mu }\right) =\frac{1}{r^{2}e^{\lambda }e^{\phi }}\partial
_{T}\left( r^{2}e^{\lambda }n\right) =0,
\end{equation}
so that the density $n\,$of the conserved quantity, say the baryon number,
is given by 
\begin{equation}
n=\frac{\rho }{e^{\lambda }r^{2}},  \label{n}
\end{equation}
where $\rho =\rho (\xi )$ should be determined by the initial condition. In
this choice of the metric, the matter Lagrangian density is expressed as 
\begin{equation}
{\cal L}_{M}=-\sqrt{-g}\varepsilon (n)=-e^{\phi }e^{\lambda
}r^{2}\varepsilon (n),  \label{LMsp}
\end{equation}
where $n$ is given by Eq.(\ref{n}). No terms with Lagrangian multipliers
appear, because the constraints are automatically satisfied. The
gravitational part is calculated as 
\begin{equation}
{\cal L}_{G}{\cal \;}=\frac{1}{2\kappa }e^{\lambda }e^{\phi }\left[
r^{\prime }\left( 2r\phi ^{\prime }+r^{\prime }\right) e^{-2\lambda }-\dot{r}%
\left( 2r\dot{\lambda}+\dot{r}\right) e^{-2\phi }+1\right] ,  \label{LGsp}
\end{equation}
where we introduced the notation $\dot{f}=\partial f/\partial T$ and $%
f^{\prime }=\partial f/\partial \xi $. In the above, we omitted the part
which can be written as the total derivative of a function, since this does
not alter the equation of motion. The total Lagrangian of the spherically
symmetric system is then given explicitly as 
\begin{equation}
{\cal L}\left[ \phi ,\lambda ,r\right] =e^{\phi }e^{\lambda }r^{2}\left\{
-\varepsilon (n)+\frac{1}{2\kappa r^{2}}\left[ r^{\prime }\left( 2r\phi
^{\prime }+r^{\prime }\right) e^{-2\lambda }-\dot{r}\left( 2r\dot{\lambda}+%
\dot{r}\right) e^{-2\phi }+1\right] \right\} .  \label{TotA}
\end{equation}
When the variation for functions $\phi ,\lambda ,$ and $r$ are in fact
arbitrary, this Lagrangian is equivalent to Einstein's equations. To see
this, we write the Euler-Lagrange equations of motion for $\phi ,\lambda ,$
and $r$ from this Lagrangian to get 
\begin{equation}
\varepsilon =\frac{1}{\kappa }\left[ \frac{1}{r^{2}}+2e^{-2\lambda }\left( 
\frac{r^{\prime }}{r}\lambda ^{\prime }-\frac{r^{\prime \prime }}{r}-\frac{%
r^{\prime \;2}}{2r^{2}}\right) +2e^{-2\phi }\left( \frac{\dot{r}^{2}}{2r^{2}}%
+\frac{\dot{r}}{r}\dot{\lambda}\right) \right] ,  \label{G00e}
\end{equation}
\begin{equation}
P=-\frac{1}{\kappa }\left[ \frac{1}{r^{2}}-2e^{-2\lambda }\left( \frac{%
r^{\prime }}{r}\phi ^{\prime }+\frac{r^{\prime \;2}}{2r^{2}}\right)
+2e^{-2\phi }\left( \frac{\ddot{r}}{r}+\frac{\dot{r}^{2}}{2r^{2}}-\frac{\dot{%
r}}{r}\dot{\phi}\right) \right] ,  \label{G11p}
\end{equation}
and 
\begin{eqnarray}
P &=&-\frac{1}{\kappa }\left[ -e^{-2\lambda }\left( \phi ^{\prime \prime
}+\phi ^{\prime 2}-\phi ^{\prime }\lambda ^{\prime }+\frac{1}{r}\left(
r^{\prime \prime }+\phi ^{\prime }r^{\prime }-\lambda ^{\prime }r^{\prime
}\right) \right) \right.  \label{G22p} \\
&&\left. +e^{2\phi }\left( \ddot{\lambda}+\dot{\lambda}^{2}-\dot{\phi}\dot{%
\lambda}+\frac{1}{r}\left( \ddot{r}+\dot{r}\dot{\lambda}-\dot{\phi}\dot{r}%
\right) \right) \right] .  \nonumber
\end{eqnarray}
We verify directly that these three equations are exactly those
corresponding to the diagonal part of Einstein's equation. In fact, writing
these equations in the form 
\begin{equation}
\varepsilon =\frac{1}{\kappa }{\cal G}_{\;0}^{0},  \label{G00}
\end{equation}
\begin{equation}
P=-\frac{1}{\kappa }{\cal G}_{\;1}^{1},  \label{G11}
\end{equation}
and 
\begin{equation}
P=-\frac{1}{\kappa }{\cal G}_{\;2}^{2},  \label{G22}
\end{equation}
where ${\cal G}_{\;0}^{0}$ ${\cal G}_{\;1}^{1}$ and ${\cal G}_{\;2}^{2}$'s
are defined, respectively, as the quantities in the square bracket $\left[
\;\;\right] $ of Eqs.(\ref{G00e}), (\ref{G11p}), and (\ref{G22p}), we can
identify the functions ${\cal G}$ as the diagonal components of the Einstein
tensor corresponding to the metric (\ref{metric}).

The only difference between our formalism here and Einstein's equation is
that in the former there is no equation corresponding to the non-diagonal
element, ${\cal G}_{\;1}^{0}$ in the latter. In Einstein's theory, this
quantity should be zero, 
\begin{equation}
{\cal G}_{\;1}^{0}=\frac{2e^{-2\lambda }}{r}\left( \dot{r}^{\prime }-\dot{r}%
\phi ^{\prime }-\dot{\lambda}r^{\prime }\right) =0.  \label{defG01}
\end{equation}
This is because in the comoving frame the energy-momentum tensor\thinspace $%
T_{\;\nu }^{\mu }$ is diagonal. Therefore, to prove that our result is
identical to the usual theory, we have to show that Eq.(\ref{defG01}) is a
consequence of Eqs.(\ref{G00}) -- (\ref{G22}). Although this proof is rather
basic matter and could be found in text books, we show it explicitly for the
sake of later discussion. We first start with the well-known Bianchi
identity (for example, see \cite{Weinberg}, p.363), 
\begin{equation}
{\cal G}_{\;\nu }^{\mu }\,_{;\mu }=\partial _{\mu }\left( \sqrt{-g}{\cal G}%
_{\;\nu }^{\mu }\right) +\frac{1}{2}\sqrt{-g}\left[ {\cal G}_{\alpha
\,\,\beta }\partial _{\nu }g^{\alpha \beta }\right] =0.  \label{Bianchi}
\end{equation}
In our case, the first component $\nu =0$ leads to 
\begin{equation}
\partial _{0}\left( \sqrt{-g}{\cal G}_{\;0}^{0}\right) +\partial _{1}\left( 
\sqrt{-g}{\cal G}_{\;0}^{1}\right) +\frac{1}{2}\sqrt{-g}\left[ \sum_{\alpha
=0}^{3}g_{\alpha \alpha }{\cal G}_{\,\,\,\alpha }^{\alpha }\partial
_{0}g^{\alpha \alpha }\right] =0.  \label{Bian1}
\end{equation}
On the other hand, we have 
\begin{eqnarray}
&&\partial _{0}\left( \sqrt{-g}{\cal G}_{\;0}^{0}\right) +\frac{1}{2}\sqrt{-g%
}\left[ \sum_{\alpha =0}^{3}g_{\alpha \alpha }{\cal G}_{\;\;\alpha }^{\alpha
}\partial _{0}g^{\alpha \alpha }\right]  \nonumber \\
&=&r^{2}e^{\phi }e^{\kappa }\left\{ \frac{\partial {\cal G}_{\;0}^{0}}{%
\partial t}+2\frac{\dot{r}}{r}\left( {\cal G}_{\;2}^{2}-{\cal G}%
_{\;1}^{1}\right) +\left( {\cal G}_{\;0}^{0}+{\cal G}_{\;1}^{1}\right)
\left( 2\frac{\dot{r}}{r}+\dot{\lambda}\right) \right\}  \nonumber \\
&=&r^{2}e^{\phi }e^{\kappa }\left[ \dot{\varepsilon}+\left( \varepsilon
+p\right) \left( 2\frac{\dot{r}}{r}+\dot{\lambda}\right) \right] =0,
\label{Bian2}
\end{eqnarray}
where Eqs.(\ref{G00}) -- (\ref{G22}) together with the energy conservation, 
\[
\dot{\varepsilon}=\frac{d\varepsilon }{dn}\frac{\partial \left( \rho
/r^{2}e^{\lambda }\right) }{\partial T}=-\left( \varepsilon +P\right) \left(
2\frac{\dot{r}}{r}+\dot{\lambda}\right) , 
\]
are used. Comparing Eqs.(\ref{Bian1}) and (\ref{Bian2}), we get 
\begin{equation}
\partial _{1}\left( \sqrt{-g}{\cal G}_{\;0}^{1}\right) =0,  \label{dG01}
\end{equation}
or 
\begin{equation}
r^{2}e^{\phi }e^{\lambda }{\cal G}_{\;0}^{1}=C(T).
\end{equation}
where $C$ is a function of $T$ only. For a non-singular metric we should
have $r(\xi =0,T)=0$, hence we conclude that $C(T)\equiv 0.$ Therefore, we
obtain 
\begin{equation}
{\cal G}_{\;0}^{1}=0.  \label{G010}
\end{equation}
This completes the proof that our result is equivalent to Einstein's
equation. That is, the Lagrangian density Eq.(\ref{TotA}) describes
correctly the dynamics of a spherically symmetric system of an ideal fluid
and gravitational field.

\subsection{Misner-Sharp Equation}

Together with Eq.(\ref{defG01}), Eq.(\ref{G010}) implies the following
relation, 
\begin{equation}
\dot{r}^{\prime }-\dot{r}\phi ^{\prime }-\dot{\lambda}r^{\prime }=0,
\label{lp}
\end{equation}
which can be obtained from Eqs.(\ref{G00}) -- (\ref{G22}) directly, without
referring to the Einstein tensor $G_{\;\nu }^{\mu }$ and its properties.

Following Ref.\cite{MS} we can express the equations of motion in a more
convenient form. First, putting $G_{\;1}^{0}\equiv 0$ in the second
component of the Bianchi identity, we have 
\begin{equation}
\frac{\partial {\cal G}_{\;1}^{1}}{\partial r}+2\frac{r^{\prime }}{r}\left( 
{\cal G}_{\;1}^{1}-{\cal G}_{\;2}^{2}\right) =-\left( {\cal G}_{\;0}^{0}+%
{\cal G}_{\;1}^{1}\right) \phi ^{\prime }.  \label{Bia3}
\end{equation}
Substituting Eqs.(\ref{G00}) -- (\ref{G22}), we get immediately that

\begin{equation}
P^{\prime }=-\left( \varepsilon +P\right) \phi ^{\prime },  \label{Eu}
\end{equation}
which is the Euler equation in the comoving frame. Now we introduce a
quantity $U$ defined by 
\begin{equation}
U=\dot{r}e^{-\phi }=\frac{dr}{d\tau },  \label{Rp}
\end{equation}
where $\tau $ is the local proper time and $d/d\tau $ is the total
derivative. The relation (\ref{lp}) is expressed in terms of $U$ as 
\[
e^{-\phi }\dot{\lambda}=\frac{U^{\prime }}{r^{\prime }}. 
\]
Now Eq.(\ref{G00e}) becomes 
\begin{equation}
8\pi G\varepsilon r^{2}=1-rr^{\prime }\left( e^{-2\lambda }\right) ^{\prime
}-e^{-2\lambda }\left( r^{\prime \prime }R+r^{\prime 2}\right)
+U^{2}+2rUU^{\prime }r^{\prime -1},  \label{e}
\end{equation}
which can be integrated as 
\begin{equation}
e^{-2\lambda }=\frac{1}{r^{\prime \;2}}\left( 1+U^{2}-\frac{2MG}{r}\right) ,
\label{elambda}
\end{equation}
where, as before, $G$ is the gravitational constant and 
\begin{equation}
M(\xi ,T)=4\pi \int_{0}^{\xi }\varepsilon r^{2}r^{\prime }d\xi =4\pi
\int_{0}^{r}\varepsilon r^{2}dr.  \label{Mt}
\end{equation}
Inserting Eq.(\ref{elambda}) into (\ref{G11}), together with (\ref{Eu}), and
after some manipulations we obtain: 
\begin{equation}
\frac{d^{2}r}{d\tau ^{2}}=\dot{U}e^{-\phi }=-\left( \frac{1}{\varepsilon +P}%
\right) \left( 1+U^{2}-\frac{2MG}{r}\right) \left( \frac{\partial P}{%
\partial r}\right) _{T}-G\frac{M+4\pi R^{3}P}{r^{2}}.  \label{MS}
\end{equation}
This form of the equation of motion was first obtained by Misner and Sharp 
\cite{MS}. Equations (\ref{Eu}),(\ref{Rp}), (\ref{Mt}) and (\ref{MS})
together with an equation of state completely determine the dynamics of a
spherical collapse, or bounce, which might be relevant for the study of the
gamma ray bursts\cite{GB}.

To see the relation between this expression and the special relativistic
Euler equation Eq.(\ref{eqm1}), we set $G=0$ in Eq.(\ref{MS}), 
\[
\frac{d^{2}r}{d\tau ^{2}}=-\left( \frac{1}{\varepsilon +P}\right) \left(
1+U^{2}\right) \left( \frac{\partial P}{\partial r}\right) _{T}. 
\]
Let $(t,r)$ the coordinate system fixed in the Minkowskian space-time. Thus,
the line element is given as 
\[
ds^{2}=dt^{2}-dr^{2}-r^{2}d\Omega ^{2}, 
\]
in a space-fixed coordinate system. However, in the comoving coordinate
system, we need non-trivial metric functions as follows. First we introduce
a coordinate transformation, 
\begin{eqnarray}
r &=&r(T,\xi ),  \nonumber \\
t &=&t(T,\xi ),  \label{transform}
\end{eqnarray}
where we may identify the coordinate $\xi $ as the comoving Lagrangian
coordinate. Thus, by definition, 
\begin{equation}
v=\left( \frac{\partial r}{\partial t}\right) _{\xi =cont.}.  \label{vel}
\end{equation}
It is always possible to choose the variable $T$ so that the cross term in
the above equation vanishes and the line element can be written in the form
of Eq.(\ref{metric}), 
\begin{equation}
ds^{2}=e^{2\phi }dT^{2}-e^{2\lambda }d\xi ^{2}-r^{2}(\xi ,t)d\Omega ^{2}.
\end{equation}
Here, 
\begin{equation}
d\tau =e^{\phi }dT|_{d\xi =0},
\end{equation}
is the (local) proper time. Note that a local Lorentz transformation relates
the infinitesimal coordinate differences to those of proper time $d\tau \;$%
and local radial distance $e^{\lambda }d\xi $ by 
\begin{equation}
\left( 
\begin{array}{c}
dt \\ 
dr
\end{array}
\right) =\left( 
\begin{array}{ll}
\gamma & v\gamma \\ 
v\gamma & \gamma
\end{array}
\right) \left( 
\begin{array}{c}
d\tau \\ 
e^{\lambda }d\xi
\end{array}
\right) .  \label{lorentz}
\end{equation}
On the other hand, since 
\begin{equation}
ds^{2}|_{d\xi =0}=d\tau ^{2}=\left[ dt^{2}-dr^{2}\right] _{d\xi
=0}=dt|_{d\xi =0}^{2}\left( 1-v^{2}\right) ,
\end{equation}
we conclude 
\begin{equation}
\gamma ^{-1}dt|_{d\xi =0}=d\tau =e^{\phi }dT.  \label{rel-1}
\end{equation}
In this way we have 
\begin{eqnarray}
\frac{d^{2}r}{d\tau ^{2}} &\equiv &e^{-\phi }\frac{\partial }{\partial T}%
\left( e^{-\phi }\frac{\partial r}{\partial T}\right) _{\xi }=\gamma \frac{%
\partial }{\partial t}\left( \gamma \frac{\partial r}{\partial t}\right)
_{\xi }  \nonumber \\
&=&\gamma \left( \frac{\partial \gamma }{\partial t}v+\gamma \left( \frac{%
\partial v}{\partial t}\right) _{\xi }\right)  \nonumber \\
&=&\gamma ^{2}\left( 1+\gamma ^{2}v^{2}\right) \left( \frac{\partial v}{%
\partial t}\right) _{\xi }.
\end{eqnarray}
Now, from Eq.(\ref{lorentz}) we have 
\begin{eqnarray*}
dt|_{dT=0} &=&v\gamma e^{\lambda }d\xi , \\
dr|_{dT=0} &=&\gamma e^{\lambda }d\xi ,
\end{eqnarray*}
so that

\begin{eqnarray}
\left( \frac{\partial P}{\partial R}\right) _{T} &=&\left( \frac{\partial P}{%
\partial r}\right) _{t}+\left( \frac{\partial P}{\partial t}\right)
_{r}\left( \frac{\partial t}{\partial r}\right) _{T}  \nonumber \\
&=&\left( \frac{\partial P}{\partial R}\right) _{t}+v\left( \frac{\partial P%
}{\partial t}\right) _{r}\;.
\end{eqnarray}
We also have 
\begin{equation}
\left( \frac{\partial v}{\partial t}\right) _{\xi }=\left( \frac{\partial v}{%
\partial t}\right) _{r}+\left( \frac{\partial v}{\partial r}\right)
_{t}\left( \frac{\partial r}{\partial t}\right) _{\xi }=\left( \frac{%
\partial v}{\partial t}\right) _{r}+v\left( \frac{\partial v}{\partial r}%
\right) _{t}.
\end{equation}
Therefore, Eq.(\ref{MS}) becomes 
\[
\left( \frac{\partial v}{\partial t}\right) _{r}+v\left( \frac{\partial v}{%
\partial r}\right) _{t}=-\frac{1}{\varepsilon +P}\frac{1}{\gamma ^{2}}\left[
\left( \frac{\partial P}{\partial r}\right) _{t}+v\left( \frac{\partial P}{%
\partial t}\right) _{r}\right] 
\]
which is exactly the relativistic Euler equation (\ref{eqm1}).

\subsection{Variational Principle in Comoving Coordinate for No Gravity
Limit ($G\rightarrow 0$)}

The above discussion suggests the possible use of the comoving (Lagrangian)
coordinate system even for cases with no gravity, that is $G\rightarrow 0$.
The effective Lagrangian presented in Sec.III is based on an ansatz for the
solution of the continuity equation in the space-fixed coordinate system. By
using a comoving Lagrange coordinate system, we may better choose the trial
function on physical grounds. Of course, the two systems of coordinates, in
principle, should be equivalent if the ansatz has enough flexibility to the
express any arbitrary flow pattern of the matter. However, for practical
applications, the appropriate choice of the coordinate system is essential
to get better results. For example, it is technically difficult to introduce
the velocity dependence in the ansatz for the density profile consistent
with the continuity equation. Therefore, for an ansatz like Eq.(\ref{nparam}%
), established in the space-fixed coordinate system, the relativistic
kinematical effects may induce some spurious effects on the dynamics of the
parameters. On the other hand, if we can choose the parametrization in the
comoving coordinate system, such kinematical effects are expected to be
automatically included in the equation of motion.

In the limit of $G\rightarrow 0$, the space-time reduces to that of
Minkowski and obviously the gravitational part of the Lagrangian density (%
\ref{TotA}) vanishes. However, in the comoving frame the line element has
still the form (\ref{metric}) and functions $e^{\phi }$ and $e^{\lambda }$
remain unknown. If we drop out the gravitational part from the action, the
variational principle does not give information on these functions. What
should be done in this limit is that Eqs.(\ref{Eu}) and (\ref{elambda}) are
used as constraints among the unknown functions, $r,\phi $ and $\lambda $.
Setting $G=0$ in Eq.(\ref{elambda}) and using the relation 
\[
\sqrt{1+U^{2}}=\sqrt{1+\left( \dot{r}e^{-\phi }\right) ^{2}}=\sqrt{1+\gamma
^{2}v^{2}}=\gamma , 
\]
we obtain 
\begin{equation}
e^{\lambda }=\frac{r^{\prime }}{\gamma }.
\end{equation}
On the other hand, assuming the isentropic initial condition and adiabatic
process, we can integrate Eq.(\ref{Eu}) with respect to $\xi $ to get 
\begin{equation}
e^{-\phi }=\frac{\varepsilon +P}{mn}\equiv h(n)  \label{efai}
\end{equation}
where $h$ is the specific enthalpy of the matter and $m$ is the rest mass of
the constituent particles. This integration constant was chosen so that $%
e^{\phi }\rightarrow 1,\;n\rightarrow 0.$

Now the action becomes 
\begin{equation}
I=-\int dT\int d\xi e^{\phi }e^{\lambda }r^{2}\varepsilon (n)=-\int dT\int
d\xi \frac{r^{\prime }r^{2}}{\gamma }\frac{mn}{\varepsilon +P}\varepsilon
(n),
\end{equation}
where 
\begin{equation}
n=\frac{\gamma \rho (\xi )}{r^{\prime }r^{2}},  \label{number}
\end{equation}
and $\rho =\rho (\xi )$ is determined from the initial condition.

Let the $T$-dependence of $r$ be specified as 
\begin{equation}
r(\xi ,t)=f(\xi ,a(T)).  \label{ansatz}
\end{equation}
Then the Lorentz factor $\gamma $ is expressed as 
\[
\gamma =\sqrt{1+e^{-2\phi }\dot{r}^{2}}=\sqrt{1+h^{2}(n)\left( \frac{%
\partial f}{\partial a}\right) ^{2}\dot{a}^{2}}, 
\]
so that the number density $n=n(a,\dot{a};\xi )$ should be determined by the
equation 
\begin{equation}
f^{\;\prime }f^{\;2}n=\rho (\xi )\sqrt{1+h^{2}(n)\left( \frac{\partial f}{%
\partial a}\right) ^{2}\dot{a}^{2}}.  \label{eqn}
\end{equation}
Finally the effective Lagrangian for $a=a(T)$ is given by 
\begin{equation}
L(a,\dot{a})=-\int d\xi \;\rho (\xi )\varepsilon \left( n\right) .
\label{eff-lag}
\end{equation}
The Euler-Lagrange equation of motion for $a$ then takes the form 
\begin{equation}
\frac{d}{dt}\int d\xi \rho (\xi )(\varepsilon +P)\frac{1}{n}\left( \frac{%
\partial n}{\partial \dot{a}}\right) =\int d\xi \rho (\xi )(\varepsilon +P)%
\frac{1}{n}\left( \frac{\partial n}{\partial a}\right) .
\end{equation}

Differently from the case of the space-fixed coordinate system, we need the
equation of state to determine the density of the matter as a function of
our dynamical variable, (\ref{ansatz}). Note the difference between the
density profiles $n(r ,t)$ of the Sec.III and $n(\xi ,T)$ defined by Eq.(\ref
{eqn}). The former is defined for constant time, $t=const$. of the space
fixed global coordinate system, and the latter is defined for constant time
coordinate, $T=const$. of the comoving coordinate system.

Although the comoving coordinate system seems physically more advantageous
than the space fixed global coordinate system in choosing an ansatz, it may
generate a difficulty in solving Eq.(\ref{eqn}) for a given equation of
state. For the ideal gas, like 
\[
P\propto n^{\gamma } 
\]
with $\gamma =5/3$ or $4/3$, an analytic solution of Eq.(\ref{eqn}) for $n$
can be obtained explicitly. However, for general cases, analytic solution is
not available. To avoid this, one may be tempted to use the proper time
defined by 
\begin{equation}
d\tau =e^{\phi }dT
\end{equation}
instead of the time coordinate $T$ and introduce the ansatz, 
\begin{equation}
r(\xi ,\tau )=f(\xi ,a(\tau )),
\end{equation}
in substitution for Eq.(\ref{ansatz}). In this case, the Lorentz factor
becomes 
\begin{equation}
\gamma =\sqrt{1+\left( \frac{\partial f}{\partial a}\right) ^{2}\left( \frac{%
da} {d\tau }\right) ^{2}}
\end{equation}
involving no density dependent term in it. Thus the density is expressed
directly as 
\begin{equation}
n=\frac{\rho (\xi )}{f^{\;\prime }f^{\;2}}\sqrt{1+\left( \frac{\partial f}{%
\partial a}\right) ^{2}\left( \frac{da}{d\tau }\right) ^{2}}
\end{equation}
without need for solving the equation, (\ref{eqn}). However, unfortunately,
the pair of variables, $\left( \tau ,\xi \right) $ do not constitute the
proper integrable coordinate system, so that the boundary condition for the
variation principle on the action, 
\begin{equation}
I=-\int d\tau \int d\xi e^{\lambda }r^{2}\varepsilon (n)  \label{action-f}
\end{equation}
is not properly defined and a simple Euler-Lagrange equation for fixed $\tau 
$ leads to a wrong result.

\section{Discussion and Concluding Remarks}

The variational approach for systems of fields, including the general theory
of relativity, is of course a basic and standard theoretical framework and
has been well studied, even for the application to the hydrodynamics
discussed here. However, to the authors' knowledge, except for the formal
derivation, no explicit variational formulation for the practical
application for relativistic hydrodynamical systems has ever been carried
out.

From the formal point of view, matter described by hydrodynamics is rather a
phenomenological concept than the consideration of the fundamental degrees
of freedom. In field theories, the variational approach is indispensable in
discussing, for example, the underlying symmetries of the matter field, such
as Noether's theorem, the quantization procedure, etc. Most of these formal
aspects of the variational approach will not be much useful for
hydrodynamical systems, except for the obvious symmetries required for the
energy-momentum tensor. Thus one might find no point in discussing
hydrodynamics from the action principle, once the equations of motion of
hydrodynamics are well established in terms of the equations for the
energy-momentum tensor.

On the other hand, as is well-known, the variational approach has practical
advantages besides its formal side. Once the variational principle is
established, we can use the method to obtain the optimal parameters of a
given family of trial solutions.

In this paper, we derived the equations of motion of hydrodynamics starting
from a very simple Lagrangian density. There it is seen that the roles of
the continuity equation as a dynamical constraint and of the local
thermodynamical relations are essential to arrive at the standard result of
hydrodynamics. When the continuity equation is soluble, such a formulation
in terms of the variational principle offers a powerful tool to obtain
approximate solutions. For a system with a high degree of symmetry such as
spherically symmetric system, we can establish the effective Lagrangian for
the density profile function. Such an effective Lagrangian is quite useful
for obtaining the approximate solutions for the hydrodynamical equation of
motion in a simple manner. Even for the finite element discretisation of the
hydrodynamic equation designed to a larger numerical solution, the
variational approach may offer a physically optimized equation of motion,
avoiding the mathematical instability with a relatively small number of
degrees of freedom\cite{Leff}. As an example of extreme simplified case, we
apply the effective Lagrangian formulation for a gas bubble in a fluid and
for the first time the relativistic version of the Rayleigh-Plesset equation
is obtained. Such an approach will be useful for the analysis of the
relativistic motions of blast waves in the models of gamma ray bursters\cite
{GB}, or the hot and dense droplet of QGP plasma, possibly formed in
high-energy nuclear collisions. The application of our formalism for such
processes is being planned. We have discussed also the introduction of
pseudo-viscosity due to Neumann and Ritchmyer in the context of variational
formulation. This will allow, for example, not only to treat
relativistically the propagation of shock waves but also to introduce the
finite relaxation time of turbulent flows in a phenomenological manner\cite
{Retard} in the relativistic fluid dynamics.

Our formalism will be useful in studying some problems of General
Relativity, too. For a spherically symmetric system, a very simple
Lagrangian density has been found. From this Lagrangian density we can show
that all the known equations of the spherically symmetric system can be
derived. We expect that, together with a good parametrization of the metric
functions, approximate solutions to these otherwise difficult problems of
stellar collapse or explosion with realistic equations of state can be
obtained. Work on this line is in progress.

Authors wish to express their thanks to Drs. C.E.Aguiar and M.O.Calv\~{a}o
for enlightening discussion and critical comments. This work is supported in
part by PRONEX (contract no. 41.96.0886.00), FAPESP(contract no. 98/2249-4)
and CNPq-Brasil Processes, 300962/86-0, 573846/1997-9, 142338/97-4 and also
by US-Department of Energy under Grant No. DE-FG03-95 ER40937, by NSF under
grant INT-9602920.

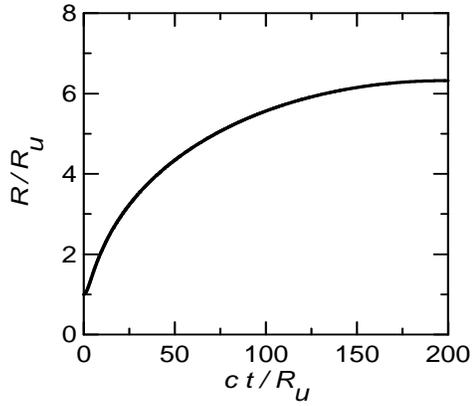
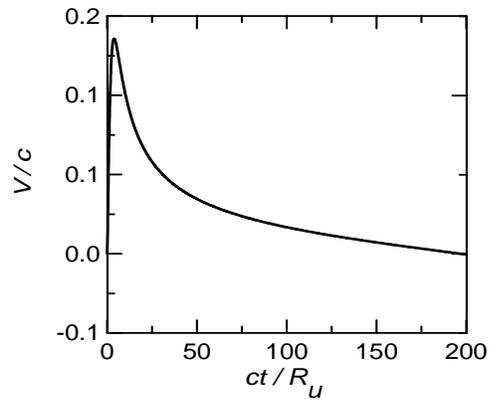
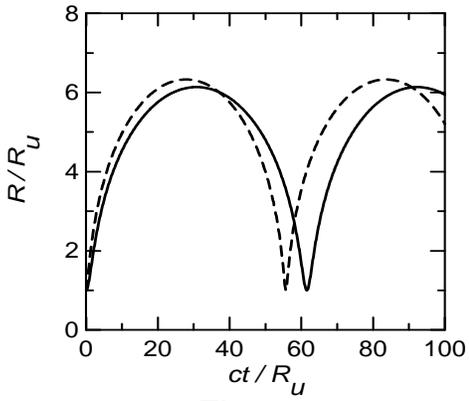
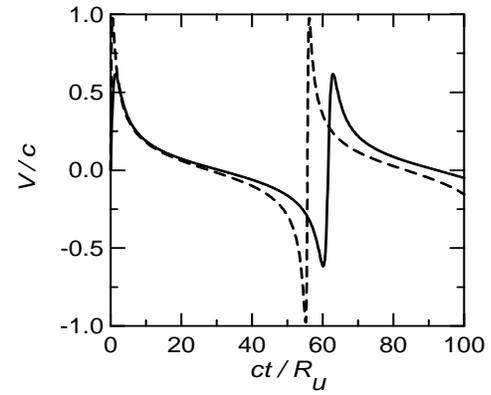
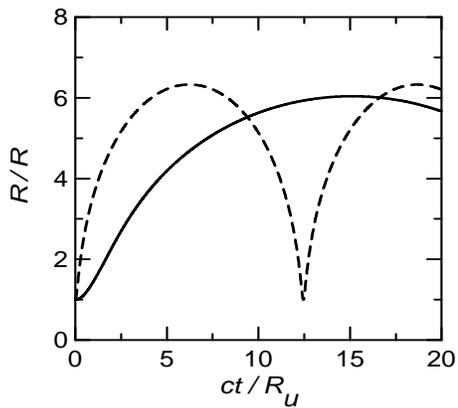
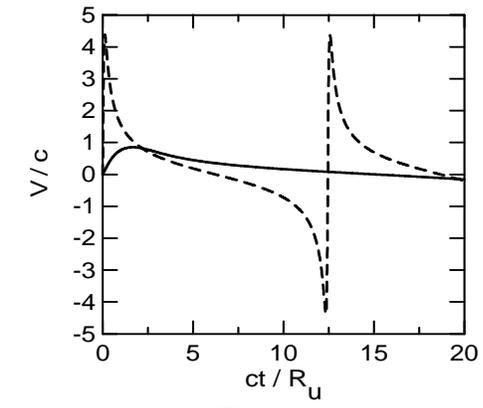
\end{document}